\documentclass[aps,showpacs,preprint,prl,superscriptaddress,endfloats*]{revtex4}

\usepackage{graphicx}
\usepackage{epsfig}

\begin{document}

\title{Strength distributions and symmetry breaking in coupled microwave billiards}

\author{B.~Dietz}
\affiliation{Institut f{\"u}r Kernphysik, Technische Universit{\"a}t
Darmstadt, D-64289 Darmstadt, Germany}
\author{T.~Guhr}
\affiliation{Matematisk Fysik, LTH, Lunds Universitet, S-22100 Lund, Sweden}
\author{H.L.~Harney}
\affiliation{Max-Planck-Institut f{\"u}r Kernphysik, D-69029
Heidelberg, Germany}
\author{A.~Richter}
\email{richter@ikp.tu-darmstadt.de}
\affiliation{Institut f{\"u}r
Kernphysik, Technische Universit{\"a}t Darmstadt, D-64289 Darmstadt,
Germany}

\date{\today}

\begin{abstract}
Flat microwave cavities can be used to experimentally simulate quantum
mechanical systems. By coupling two such cavities, we study the
equivalent to the symmetry breaking in quantum mechanics. As the coupling is
tunable, we can measure resonance strength distributions as a function
of the symmetry breaking. We analyze the data employing a qualitative
model based on Random Matrix Theory (RMT) and show that the results
derived from the strength distribution are consistent with those
previously obtained from spectral statistics.
\end{abstract}


\maketitle

The breaking of quantum mechanical symmetries represents a prominent
object of study in a rich variety of systems, ranging from high energy
to condensed matter physics. Sometimes it is possible to determine the
size of the symmetry breaking by analyzing statistical observables. We
mention parity violation~\cite{BowmanMBW}, the breaking of atomic and
molecular symmetries~\cite{RosenzweigPorter,Haller} and isospin
mixing~\cite{Aberg,HRW,Mitchell,GuhrWeidenm,Shriner,BGHTheo} in
nuclei.  Symmetry breaking influences the spectral statistics as well
as the wave function statistics, such as the distributions of partial
widths and transition matrix elements. While symmetry breaking cannot
be controlled or tuned in nuclei, atoms and molecules, this was
possible for elastomechanical resonances in quartz crystals by
successively breaking crystal symmetries~\cite{quartz}. An analogy to
the nuclear case of symmetry breaking was given by two coupled
microwave cavities~\cite{gekop} with a tunable coupling. The
statistical properties of both systems are fully equivalent to those
of a quantum system~\cite{Stockmann,IMAPRO}. Wheras these two
investigations focused on the spectral statistics, width distributions
for two different types of resonances in elastic aluminum plates were
studied in Ref.~\cite{Anders} for a varying degree of mixing.

Here, we present experimental results on the distribution of products
of partial widths in two coupled chaotic microwave cavities. First, we
measure the distribution for different couplings. It can be normalized
such that it only depends on the symmetry breaking. As this is
different from Ref.~\cite{Anders}, we can apply a
qualitative statistical model which extends the model of 
Ref.~\cite{BGHTheo} in order to,
second, extract the size of the symmetry breaking 
from the data. We then show, third, that the symmetry
breaking thereby obtained is consistent with the one previously found
from the spectral statistics.

\begin{figure}
\begin{center}
\epsfig{figure=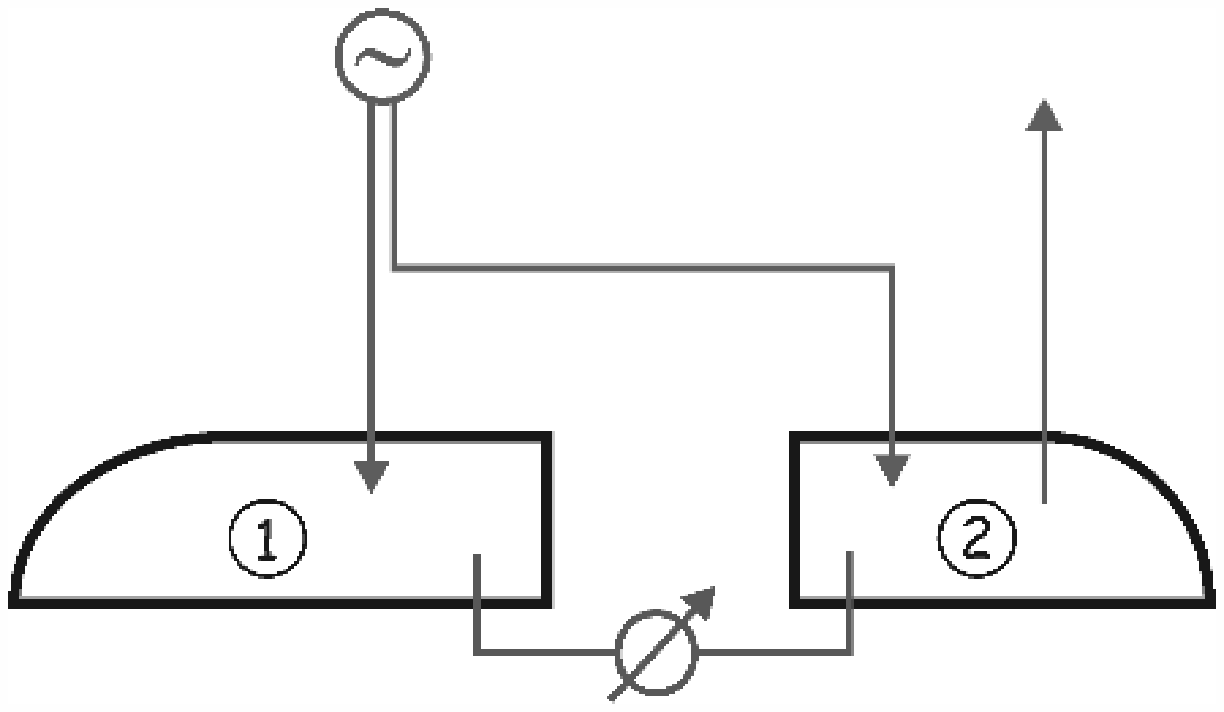,width=8.5cm,angle=0}
\caption{\label{billgekop}
The two cavities are coupled with a tunable coupling}
\end{center}
\end{figure}

In the experiment, we used two flat cylindrical microwave cavities
having the shape of the quarter of a Bunimovich stadium\cite{gekop}. 
In both resonators (see Fig.~\ref{billgekop}) the radius of the
quarter circle is $0.2$~m. The ratios of the length of the rectangular
part to the radius are $\gamma_1=1$ and $\gamma_2=1.8$,
respectively. Therefore the level densities increase with different
slopes as functions of frequency ~\cite{Weyl,Baltes}. The cavities
were put on top of each other and circular holes, 4~mm in diameter,
were drilled through the walls of both resonators. The coupling was
realized by a niobium pin, 2~mm in diameter, penetrating through the
holes into both resonators. Due to the ring-shaped gap between the
niobium pin and the hole surface, the pin acts like an antenna, which
in the experimental frequency range supports one TEM-mode. The
coupling is controlled by the penetration depth~\cite{gekop}.

\begin{figure}
\begin{center}
\epsfig{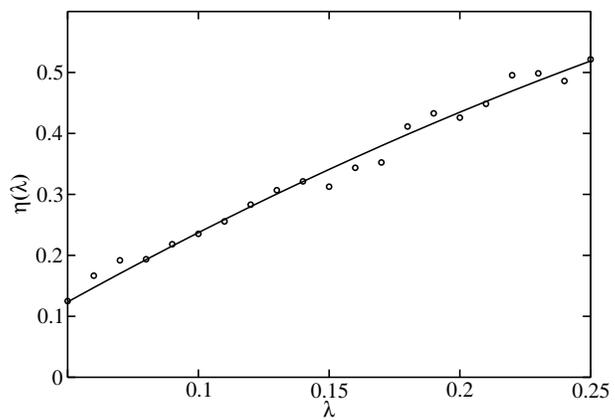}
\caption{\label{FigEta}
Size of the tunable coupling $\lambda$ between the two cavities
sketched in Fig. \ref{billgekop}.  The function $\eta(\lambda)$ enters
Eq. (\ref{eq:lam}).}
\end{center}
\end{figure}

Information about the partial widths is obtained from transmission
spectra. At a given frequency $f$, the relative power transmitted from
antenna $a$ to antenna $b$ is proportional to the absolute square of
the scattering matrix element, $P_{{\rm out},b}/P_{{\rm in},a}\sim
|S_{ab}(f)|^2$.  For sufficiently isolated resonances, one has
\begin{equation}
S_{a b}(f) = \delta_{a b} 
    - i\frac{\sqrt{\Gamma_{\mu a}\Gamma_{\mu b}}}
	     {f-f_\mu + \frac{i}{2}\Gamma_\mu}\, 
\label{SMatrix}
\end{equation}
for $f$ close to the frequency $f_{\mu}$ of the $\mu$-th
resonance. The quantities $\Gamma_{\mu a}$ and $\Gamma_{\mu b}$ are
the partial widths related to the antennae $a$ and $b$, $\Gamma_{\mu}$
is the total width of the resonance \cite{QCDWidths}. In the
experiment, three antennae were attached to each resonator, 
where half of the microwave power was fed into each resonator (see
insert of Fig.~\ref{FigEta}). 
Thereby, altogether
six transmission spectra were obtained for 4 different couplings; in
the order of increasing coupling these are denoted as (8,0), (5,3),
(4,4) and (5,8) in Ref.~\cite{gekop}. Here, the pair $(x_1,x_2)$
denotes the penetration depth in mm of the pin into either cavity.
The transmission spectra were measured up to a maximum frequency of
17.5~GHz. 
The resonance strengths $\Gamma_{\mu a}\Gamma_{\mu b}$, i.e.~the
products of the partial widths, are determined as described in
\cite{strengthdistribution}. 
Resonances with peak heights below a certain value may not be
detected, implying that some strength is missing in the tails of the
distributions \cite{strengthdistribution}. We will show that this has
no effect on the results.

We work with the resonance strengths $\Gamma_{\mu a}\Gamma_{\mu b}$
and their distributions. They yield the same information as the
distributions of the partial widths $\Gamma_{\mu a}$ themselves. The
data are {\it unfolded} as in
Refs.~\cite{QCDWidths,strengthdistribution}.

To interpret the distribution of the empirical data $\Gamma_{\mu
a}\Gamma_{\mu b}$, we employ a statistical model which 
is a special case of the Rosenzweig--Porter
model~\cite{RosenzweigPorter}.  A symmetry is associated with a
quantum number. If it suffices to consider only two different values
of it, the Hamiltonian can be written in the form
\begin{equation}
H=\pmatrix{H_1 & 0   \cr
           0   & H_2 \cr 
          }\, +
\alpha\pmatrix{0   & V   \cr
         V^T & 0   \cr
          }\, ,
\label{HamMod}
\end{equation} 
where the first part preserves the symmetry and the second part breaks
it with the parameter $\alpha$. For isospin mixing in nuclei, $H_1$
and $H_2$ correspond to the sub--Hamiltonians for two isospin quantum
numbers, while $V$ accounts for the Coulomb interaction that mixes
isospin. In our case the ``symmetry'' preserving situation is simply
realized by considering the states of each uncoupled cavity as
eigenstates of a ``symmetry operator''.  Thus, $H_1$ and $H_2$ model
the dynamics in the two cavities, without the coupling. We choose
their dimensions $N_1$ and $N_2$ different because the level densities
in the cavities have different slopes as a function of frequency.  For
later purposes, we introduce the {\it fractional densities}
$g_1=N_1/(N_1+N_2)$ and $g_2=N_2/(N_1+N_2)$.  As the whole system is
time--reversal invariant and the dynamics in the Bunimovich billiards
is fully chaotic, we represent $H_1$ and $H_2$ by real--symmetric
random matrices drawn from the Gaussian Orthogonal Ensemble (GOE)
~\cite{Mehta,GMGW,Stockmann, Haake}. The coupling is modeled by the
off--diagonal blocks $V$ and $V^T$ in Eq.~(\ref{HamMod}), where the
matrix $V$ is real with no symmetries and has dimension $N_1\times
N_2$. In the experiment the modes in one resonator are coupled to
those in the other via one TEM-mode. We model the coupling of the
$N_1$ and $N_2$ resonator modes with this TEM mode by an
$N_1$-dimensional vector $v$ and an $N_2$-dimensional vector $w$. The
coupling matrix $V$ then acquires the dyadic structure $V=vw^T$. The
matrix $V$ has rank $M=1$. The entries of $v$ and $w$ are chosen as
Gaussian random numbers. The distribution of the elements $V_{nm}$ is
\begin{equation}
q(V_{nm}\vert\sigma)~{\rm d}V_{nm}=
               \frac{K_0\left(V_{nm}/\sqrt{\sigma}\right)}{\pi\sqrt{\sigma}}
              ~{\rm d}V_{nm} \ ,
\label{eq:VertVnm}
\end{equation}
where $K_0$ is the modified Bessel function of zeroth order~\cite{GR}.
The variance $\sigma$ of the elements of $V$ is adjusted exactly as
described in ~\cite{GuhrWeidenm,BGHTheo}. The vertical bar on the
l.h.s. separates the argument of $q$ into the random variable $V_{nm}$
and the parameters, i.e.~in this case $\sigma$. Universal spectral
fluctuations have to be measured on the scale of the local mean level
spacing $D$. Accordingly, the parameter measuring the size of the
coupling or the symmetry breaking is $\lambda
=\alpha/D$~\cite{DysonPandey}. It connects to the spreading width via
the relation $\Gamma^\downarrow=2\pi \alpha^2/D= 2\pi\lambda^2
D$~\cite{HRW}. {\it Spectral fluctuations} have been calculated
perturbatively in $\lambda$ in Refs.~\cite{FKPT,LeitnerTh} and exactly
for the time--reversal non--invariant case in
Refs.~\cite{GuhrWeidenm,Pan}. For the analysis of the spectral
properties of the coupled microwave billiards
\cite{gekop}, a model Hamiltonian of the form Eq.~(\ref{HamMod}) with a coupling
matrix $V$ of rank $N_1$ was applied. While, in the relevant range of
$\lambda$-values the spectral properties do not depend on the rank of $V$,  
the statistical properties of the eigenvector components deviate 
for $M\ll N_1$.

Accordingly we apply and extend the qualitative model of
Ref.~\cite{BGHTheo} where the statistics of {\it transition} strengths
was considered. In a fully chaotic system, the partial width
distribution converges to the Porter-Thomas
form~\cite{PorterThomas,GMGW,Stockmann,Haake}
\begin{equation}
{\rm PT}(t_a\vert\tau_a){\rm d}t_a = \frac{1}{\sqrt{2\pi t_a/\tau_a}}
                       \exp\left(-{t_a\over 2\tau_a}\right)
                       {{\rm d}t_a\over\tau_a}\, 
\label{eq:PT}
\end{equation}
for a large level number. We write $t_a$ for the partial width instead
of $\Gamma_{\mu a}$, because the distribution does not depend on the
resonance index $\mu$. In practice the distribution is obtained from
the sample of all resonances.  Its first moment equals $\tau_a$. For
two coupled chaotic systems the partial width distribution
$p(t_a\vert\lambda,\tau_a)$ involves the symmetry breaking Hamiltonian
in Eq.~(\ref{HamMod}) for large level numbers. It is easily
constructed in the limiting case without symmetry breaking; i.e.~for
$\lambda =0$ we have
\begin{eqnarray}
p(t_a\vert\lambda=0,\tau_a)
 &=& g_1^2 {\rm PT}\left(t_a\bigg\vert\frac{\tau_a}{g_1}\right) +
     g_2^2 {\rm PT}\left(t_a\bigg\vert\frac{\tau_a}{g_2}\right) 
                                                     \nonumber\\
 & & \qquad    + 2g_1g_2\delta\left(\frac{t_a}{\tau_a}\right) \, .
\label{eq:lam0}
\end{eqnarray}
The distribution in Eq.~(\ref{eq:lam0}) again has the expectation
value $\tau_a$. We now model the case of small symmetry breaking
$\lambda$ by an interpolating ansatz: We expect the Porter-Thomas
distributions in Eq.~(\ref{eq:lam0}) to maintain their shape, such
that only their width parameters change. The delta function in
Eq.~(\ref{eq:lam0}) acquires a width and develops into a non-singular
function. Numerical studies have lead us to approximate it
by~\cite{FT1} $P_0(t\vert\rho)~{\rm d}t={K_0\left(\sqrt{t/\rho}\,
\right)}/
\left({\pi\sqrt{t\rho}}\right) ~{{\rm d}t}$, where $\rho$ stands for the variance parameter.
We arrive at the model
\begin{eqnarray}
p(t_a\vert\lambda,\tau_a)
 &=& g_1^2 {\rm PT}\left(t_a\vert\tau_a\kappa_1(\eta)\right) +
     g_2^2 {\rm PT}\left(t_a\vert\tau_a\kappa_2(\eta)\right)
                                                     \nonumber\\
 & & \qquad    + 2g_1g_2{\rm P_0}(t_a\vert\tau_a\eta^2) \ .
\label{eq:lam}
\end{eqnarray}
The shape of the whole distribution is determined by the quantities
$\kappa_1(\eta)$, $\kappa_2(\eta)$ and $\eta =\eta(\lambda)$ which all
depend on $\lambda$. With the help of the limiting case
Eq.~(\ref{eq:lam0}), we can come up with educated guesses for these
functions. We must have $\eta(0)=0$ and furthermore $\kappa_j(0)=1/g_j
, \ j=1,2$.  As the functions $\kappa_j$ ought to be even in $\eta$,
we choose $\kappa_j(\eta)=1/g_j+\left(1-1/g_j\right)\eta^2 \, , \quad
j=1,2$. To construct the function $\eta=\eta(\lambda)$, we fit the
ansatz given in Eq.~(\ref{eq:lam}) including these choices to Monte
Carlo simulations of the distributions involving the Hamiltonian in
Eq. (\ref{HamMod}) for $N_1+N_2=300$. We do that for 21 values of
$\lambda$ in the interval $[0.05,0.25]$ which is the relevant
parameter range in the experiment.  The fractional densities in the
experiment are $g_1=0.59$ and $g_2=0.41$. In Fig.~\ref{FigEta} we
display the values thus obtained for $\eta(\lambda)$. The functional
form is well described by the polynomial
$\eta(\lambda)=2.57\lambda-1.98\lambda^2$.

{}From the distributions for two partial widths $t_a$ and $t_b$, say,
we obtain the distribution for the resonance strength $y=t_at_b$ as in
\cite{strengthdistribution} by a filter integration.
As we know the fractional densities $g_1$ and $g_2$ in the experiment,
we arrive at a qualitative model for the resonance strength
distribution that depends only on one single parameter --- the measure
$\lambda$ for the symmetry breaking. The distributions in
Eq.~(\ref{eq:lam}) and accordingly the strength distribution diverge
for $t_a\rightarrow 0$. Thus, for their graphical representation we
use the logarithmic variable $z=\log_{10}(y/\tau_a\tau_b)$.

\begin{figure}
\begin{center}
\epsfig{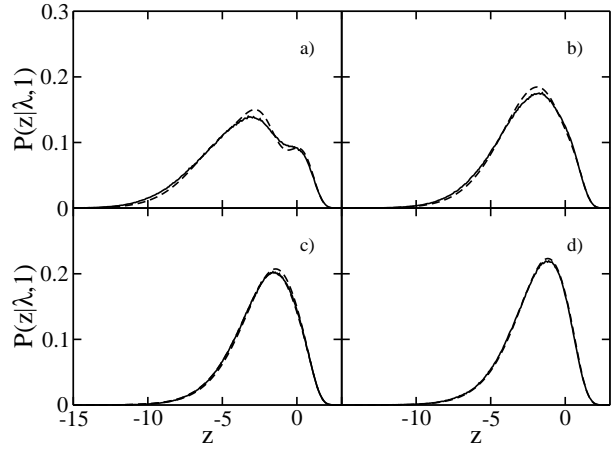}
\caption{\label{FigProductdistr1}
Monte Carlo simulation of resonance strength distributions (solid
lines) compared to the calculated strength distribution (dashed
lines) for the couplings $\lambda =0.03$ (a), $\lambda =0.10$ (b),
$\lambda =0.17$ (c), $\lambda =0.24$ (d).} 
\end{center}
\end{figure}

\begin{figure}
\begin{center}
\epsfig{figure=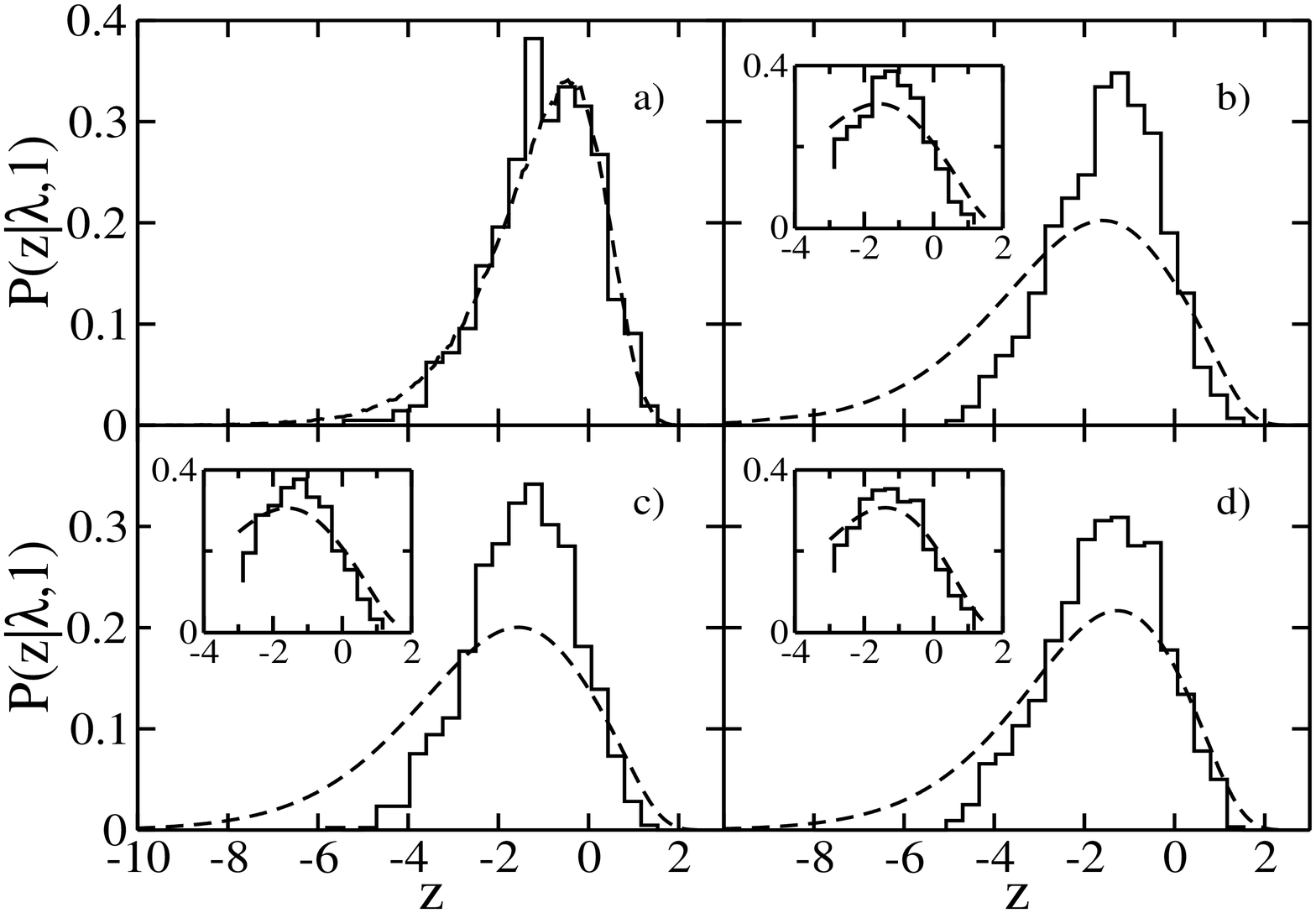,width=8cm,angle=0}
\caption{\label{FigExperiment}
The experimental resonance strength distributions
(histograms) for the couplings (8,0) (a), (5,3) (b), (4,4) (c) and
(5,8) (d) fitted to the calculated strength distribution. 
The symmetry breaking parameter $\lambda$ equals 
$\lambda =0$ in (a), $\lambda =0.110$ in (b), $\lambda =0.125$ in (c), 
and $\lambda =0.185$ in (d). The insets in b), c), and d) show experimental 
strength distributions together with RMT model fits for a $z$ interval [-3,1.5]
in which the probability of missing strength is small. }
\end{center}
\end{figure}

Despite its simplicity, the qualitative model in Eq.~(\ref{eq:lam})
yields a satisfactory description. We demonstrate this in
Fig.~\ref{FigProductdistr1} where a Monte Carlo simulation of
resonance strength distributions obtained from the random matrix
Hamiltonian in Eq.~(\ref{HamMod}) with the same fractional densities
as in the experiment is compared to the calculated strength
distribution for four different values of $\lambda$. The strongest
deviations between both curves are observed for the smallest value of
$\lambda$. For values $\lambda < 0.03$ the description by the
qualitative model ceases to be satisfactory.  But, most importantly,
the position of the maximum is described very well in all four
examples. It is particularly this feature which makes the qualitative
model in Eq.~(\ref{eq:lam}) useful for the analysis of experimental
data. The fit of the model to the experimental data is displayed in
Fig.~\ref{FigExperiment}. Due to the above mentioned missing strength
in the left tails, the shape of the distribution is not described
quantitatively. In fact, for the determination of $\lambda$ only
strengths in a $z$ interval [-3,1.5] were taken into account, where
the probability of missing strength is small. The corresponding
experimental distributions are shown together with the RMT model fits
in the insets of Fig.~\ref{FigExperiment} b)-d). For very small values
of $\lambda$ (cf. Fig.~\ref{FigExperiment} a)) the distributions agree
fairly well in this interval. We emphasize again that the peak
positions in the chosen interval of $z$ values uniquely determine the
coupling strengths. We thus expect reliable results for the parameter
$\lambda$. To carefully check how much the missing strength influences
the extracted values of $\lambda$, we amended the qualitative model by
taking care of the experimental thresholds using an analytic
ansatz~\cite{strengthdistribution}. As expected, the influence of the
thresholds on the extracted values of $\lambda$ turned out to be
negligible, except when $\lambda$ is essentially zero. As only the
position of the maximum in the strength distribution is relevant, we
have some freedom in choosing the distribution $P_0$ in
Eq.~(\ref{eq:lam}). We also tried a Porter-Thomas distribution, but
the $K_0$--one describes the shape of the resonance strength
distribution better.

We finally compare the coupling strengths $\lambda$ found in the
present work with those extracted from the {\it spectral correlations}
in Ref.~\cite{gekop,Abul}.  The values are given in Tab.~\ref{table1}.
\begin{table}
\caption{\label{table1}
Symmetry breaking parameter $\lambda$ from the spectral correlations
and from the resonance strength distributions.}
\label{table2}
\begin{tabular}{cp{1.5cm}ccccccc}
\hline 
\hline 
Physical &\   &$\lambda$   &\   &\   &\   &$\lambda$\\
coupling  &\   &Ref.~\cite{gekop} &\   &\   &\     &Present work\\
\hline
(8,0)  &\   &$\leq 0.029$     &\   &\   &\   &$< 0.003$\\
(5,3)  &\   &$0.105\pm 0.008$ &\   &\   &\   &$0.116\pm 0.003$\\
(4,4)  &\   &$0.130\pm 0.007$ &\   &\   &\   &$0.122\pm 0.004$\\
(5,8)  &\   &$0.180\pm 0.006$ &\   &\   &\   &$0.195\pm 0.007$\\ 
\hline 
\hline 
\end{tabular}
\end{table}
The $\lambda$ values obtained in the present work are averaged over
all six antennae combinations. The two analyses agree within the
experimental errors. For the weakest coupling only an upper limit can
be given. This is due to the fact that for essentially zero coupling a
considerable share of strength lies below the experimental threshold
of detection.  The consistency for the other couplings is an
encouraging corrobaration of the analysis carried out in this
contribution~\cite{consistency}.

In conclusion, we have measured resonance strength distributions for
two coupled microwave billiards modeling quantum systems with symmetry
breaking. We analyzed the data with a qualitative model which depends
only on one single free parameter --- the size of the symmetry
breaking. Our results are interesting from an additional point of
view. The spectral correlations are more strongly affected by missing
levels than partial widths and transition or resonance strength
distributions are affected by missing strength. This is so because the
latter do not comprise {\it correlations}, they are just {\it
densities}. Thus the observables related to the wave functions may
provide more reliable information than the spectral correlations. We
have shown, with data much better than are usually available, that the
empirical information extracted from the wave function observables is
consistent with that obtained from the spectral correlations.

We thank T.~Seligman for hospitality at the Centro Internacional de
Ciencias in Cuernavaca, Mexico, where Part of this work was done, and
H.J.~St\"ockmann for valuable discussions concerning the modeling of
the coupling as well a C.~Dembowski for extracting the resonance
parameters from the spectra \cite{Dembo}. This work has been supported
by the DFG within SFB 634 and by Det Svenska Vetenskapsr\aa det. One
of us (A.R.) is grateful to the latter for the Tage Erlander Guest
Professorship 2006.

%
%
%
%
%


\begin{thebibliography}{99}
 
\bibitem{BowmanMBW} J.~D.~Bowman {\it et al.},
Annu. Rev. Nucl. Part. Sci. {\bf 43}, 829 (1993);
G.E. Mitchell, J.D. Bowman, and H.A. Weidenm\"uller, Rev. Mod. Phys. 71,
445 (1999).

\bibitem{RosenzweigPorter}
N. Rosenzweig and C. E. Porter, Phys. Rev. {\bf 120}, 1698 (1960).

\bibitem{Haller}
E. Haller, H. K\"{o}ppel, and L. S. Cederbaum,
Chem. Phys. Lett. {\bf 101}, 215 (1983).

\bibitem{Mitchell}
G. E. Mitchell {\it et al.},
Phys. Rev. Lett. {\bf 61}, 1473 (1988).

\bibitem{HRW} H.~L.~Harney, A.~Richter, and H.~A.~Weidenm\"uller, Rev. Mod.
Phys. {\bf 58}, 607 (1986).

\bibitem{GuhrWeidenm} T.~Guhr and H.~A.~Weidenm\"uller, Ann. Phys. (N.Y.) 
{\bf 199}, 412 (1990).

\bibitem{Shriner}
J. F. Shriner, Jr., G. E. Mitchell, and B. A. Brown, Phys. Rev. C {\bf 71},
024313 (2005), and refs. therein.

\bibitem{BGHTheo}C.I.~Barbosa, T.~Guhr, and H.L.~Harney,
Phys.~Rev.~E \textbf{62}, 1936 (2000).

\bibitem{Aberg} S.~{\AA}berg {\it et al.},
Phys. Lett. B {\bf 598}, 42 (2004). 

\bibitem{quartz}
C.~Ellegaard {\it et al.},
Phys. Rev. Lett. {\bf 77}, 4918 (1996).

\bibitem{gekop}H. Alt {\it et al.},
Phys. Rev. Lett. {\bf 81}, 
4847 (1998).

\bibitem{Stockmann}   H.-~J. St\"{o}ckmann,
                      {\it Quantum Chaos - An Introduction},
                      (Cambridge University Press, Cambridge, 1999).

\bibitem{IMAPRO} A. Richter, in: \emph{Emerging
Applications of Number Theory, The IMA Volumes in Mathematics and its
Applications}, Vol. {\bf 109}, edited by D.A. Hejhal
{\it et al.}, p. 479, (Springer, New York, 1999).

\bibitem{Anders}
A.~Andersen {\it et al.},
Phys. Rev. E {\bf 63}, 066204 (2001).

\bibitem{Weyl} H.~Weyl, {\it J. Reine Angew. Math.} {\bf 141}, 1 (1912);
{\it ibid} 163, {\it J. Reine Angew. Math.} {\bf 143}, 177 (1913).

\bibitem{Baltes} H.~P.~Baltes and E.~R.~Hilf, {\it Spectra of Finite Systems}
(Bibliographisches Institut Mannheim, 1975).

\bibitem{QCDWidths}H. Alt {\it et al.},
Phys. Rev. Lett. \textbf{74}, 62 (1995).

\bibitem{strengthdistribution}
C.~Dembowski {\it et al.},
Phys. Rev. E {\bf 71}, 046202 (2005).

\bibitem{Mehta}       M.~L.~Mehta,
                      {\it Random Matrices}
                      (Academic Press, New York, 1991), 2nd ed.
\bibitem{GMGW}        T. Guhr, A. M\"uller--Groeling,
                      and H.A. Weidenm\"uller,
                      Phys. Rep. {\bf 299}, 189 (1998).
\bibitem{Haake}       F.~Haake,
                      {\it Quantum Signatures of Chaos},
                      2nd edition,
                      (Springer Verlag, Berlin 2001).

\bibitem{GR} I.S. Gradshteyn and I.M. Ryzhik, \emph{Tables of Integrals, 
Series, and Products.}, (Academic Press, New York, 1980).

\bibitem{DysonPandey} F.J. Dyson,
                      J. Math. phys. {\bf 3}, 1191 (1962);
                      A. Pandey,
                      Ann. Phys. (NY) {\bf 134}, 110 (1981).

\bibitem{LeitnerTh}
D. M. Leitner, Phys. Rev. E {\bf 48}, 2536 (1993); 
D. M. Leitner, H. K\"{o}ppel, and L. S. Cederbaum,
Phys. Rev. Lett. {\bf 73}, 2970 (1994).

\bibitem{FKPT}        J.B. French {\it et al.},
                      Ann. Phys. (NY) {\bf 181}, 198 (1988).                      

\bibitem{Pan}         A. Pandey, 
                      Chaos, Solitons and Fractals {\bf 5}, 1275 (1995).

\bibitem{PorterThomas}
C.E.~Porter, \emph{Statistical Theories of
Spectra: Fluctuations}, (Academic Press, New York, 1965).

\bibitem{FT1} For very weak symmetry breaking, this follows
from the $K_0$-distribution of the matrix elements $V_{nm}$.

\bibitem{Abul} A.~Y.~Abul-Magd {\it et al.},
Phys. Rev. E {\bf 65}, 056221 (2002).
 
\bibitem{consistency} However, this consistency does not prove that the 
dyadic coupling specified prior to Eq.~(\ref{eq:VertVnm}) is necessary
to describe the present data. It is required by the experimental set-up. 
We have convinced ourselves that the qualitative model in the version of 
reference \cite{BGHTheo} with high-dimensional coupling yields this
consistency, too.

\bibitem{Dembo} C.~Dembowski, Dissertation D17, TU-Darmstadt (2003).
\end{thebibliography}
\end{document}